\documentstyle[12pt]{article}

\newcommand{\resetcounter}{\setcounter{equation}{0}}     % set counter to zero

\begin{document}
%------------------------------------------------------------------------------

%-------------------------TITLEPAGE-------------------------------------
\thispagestyle{empty}
\begin{titlepage}
\begin{flushright}
HUB-IEP-95/14 \\
hep-th/9509042 \\
September 1995
\end{flushright}
\vspace{0.3cm}
\begin{center}
\Large \bf Topological Partition Function
           \\ and String-String Duality
\end{center}
\vspace{0.5cm}
\begin{center}
Gottfried \ Curio$^{\hbox{\footnotesize{1,2}}}$ ,  \\
{\sl Institut f\"ur Physik, Humboldt--Universit\"at,\\
 Invalidenstrasse 110, D--10115 Berlin, Germany}
\end{center}
\vspace{0.6cm}

\begin{abstract}
\noindent
The evidence for string/string-duality can be extended from the matching of the
vector couplings to gravitational couplings. In this note this is shown in the
rank three example, the closest stringy analog of the Seiberg/Witten-setup,
which is related to the Calabi-Yau $WP^4_{1,1,2,2,6}(12)$. I provide  an exact
analytical verification of a relation  checked by coefficient comparison to
fourth order by Kaplunovsky, Louis and Theisen.
\end{abstract}

\vspace{0.3cm}
\footnotetext[1]{E-MAIL:curio@qft2.physik.hu-berlin.de}
\footnotetext[2]{Supported by Deutsche Forschungsgemeinschaft}
\vfill
\end{titlepage}

%-----------------END OF TITLEPAGE----------------------------------------

\setcounter{page}{1}

\resetcounter
Recent progress in nonperturbative understanding of string theory is based on
the conjectured  string-string duality [\ref{D}],[\ref{W}] first proposed in
$D=6$ for the heterotic string on $T^4$ and the type IIA string on $K3$  , then
extended and partially verified in $D=4$ for  the heterotic string on $K3\times
T^2$ and type IIA on  a Calabi-Yau
[\ref{FHSV}],[\ref{KV}],[\ref{KLM}],[\ref{KKLMV}]. In the course of
accumulating evidence for the conjectured equivalence of $N=2$ string theories
in [\ref{KLTh}] new material was added in comparing the holomorphic
$F_1$-functions describing 1-loop gravitional couplings of vectormultiplets in
the proposed [\ref{KV}] dual pair consisting of the heterotic string on
$K3\times T^2$ with gauge group $U(1)^3$ (graviphoton, vector partner of the
dilaton and of the toroidal modulus T; the second toroidal modulus U is locked
at $U=T$) and the type IIA string on a suitable Calabi-Yau (cp. also
[\ref{AGNT}],[\ref{AP}]).
Concretely Kaplunovsky, Louis and Theisen [\ref{KLTh}] showed for the rank
three model of [\ref{KV}] that

%------------------------------------------------------
\begin{eqnarray*}
F_{1}^{het} =
\frac{6}{4\pi^2}\log[y^{-2}(j(t)-j(i))^{\frac{1}{6}}\eta^2(t)^{-50}]\;,
\end{eqnarray*}

 %-----------------------------------------------------

where\footnote{We use their identification of the modular invariant dilaton(cp.
also [\ref{dWKLL}]) got by matching the two prepotentials.}

%----------------------------------------------------
\begin{eqnarray*}
t &=& t_1=iT\\
y &=& e^{-8\pi^2S^{inv}} = g(iT)e^{-8\pi^2S}
\end{eqnarray*}
%----------------------------------------------------

with $t_1$ and $t_2=4\pi iS$ the special coordinates (the complexified K\"ahler
class is related to the cohomology classes of the relevant divisors (cp.
[\ref{CdlOFKM}]) by $B+iJ=t_1H+t_2L$) of $WP^4_{1,1,2, 2,6}(12)^{-252}_{2,128}$
, which are here already matched with their heterotic counterparts. The  mirror
map for the complex structure deformations of the mirror Calabi-Yau coming from

%------------------------------------------------------
\begin{eqnarray*}
p=z_1^{12}+z_2^{12}+z_3^6+z_4^6+z_5^2+a_0z_1z_2z_3z_4z_5+a_1z_1^6z_2^6
\end{eqnarray*}
%------------------------------------------------------

is given by ($q_j=e^{2\pi it_j}$)

%-----------------------------------------------------
\begin{eqnarray*}
x=j(q_1)^{-1}+O(q_2), y=g(q_1)q_2+O(q_2^2)
\end{eqnarray*}
%-----------------------------------------------------

(with uniformizing variables at large complex structure
$x=a_1a_0^{-6}$,$y=a_1^{-2}$)\footnote{The parameters of [\ref{KLTh}] and
[\ref{CdlOFKM}] are related by $a_0=-12\psi , a_1=-2\phi $.}.

 At weak coupling  the first four terms in $q_1$ are then matched [\ref{KLTh}]
by comparing to a series expansion for $F_1^{II}$.

 I will show here exact agreement in $q_1$ (at weak coupling).
Let us first reformulate the heterotic result as

%-------------------------------------------------------
\begin{eqnarray*}
F_1^{het} \sim \log[y^{-\alpha}(j(t)-j(i))^{\beta}\eta^2(t)^{-\gamma}]\;  ,
\end{eqnarray*}
%-------------------------------------------------------

where in
%-------------------------------------------------------
\begin{eqnarray*}
\alpha &=& 12\beta\\
\gamma &=& 300\beta
\end{eqnarray*}
%-------------------------------------------------------

 the 12 (the 24 in  equations (18), (22) of [\ref{KLTh}]) comes from curvature
coupling normalization and the 300 is $b_{grav}=48-\chi$ [\ref{KLTh}].

 I will show first the following on the type II side ( for some numbers
$\alpha, \beta, \gamma, c$ )

%-------------------------------------------------------
\begin{eqnarray*}
F_1^{II}=\log[ y^{-\alpha} E_4^{\frac{c-1}{2}}(t) (j(t)-j(i))^{\beta}
\eta^2(t)^{-\gamma}]
\end{eqnarray*}

%-------------------------------------------------------

(up to an additive constant) with $\gamma=2(c+3-\frac{\chi}{12})$.
 By the holomorphic anomaly of$F_1$ [\ref{BCOV}] we have [\ref{CdlOFKM}]
%-------------------------------------------------------
\begin{eqnarray*}
F_1^{II}=\log[(\frac{12\psi}{\omega _0})^{5-\frac{\chi}{12}}
\frac{\partial(\psi ,\chi )}{\partial (t_1 ,t_2)} f ]
\end{eqnarray*}
%--------------------------------------------------------

with $f=\Delta^a (\phi^2 -1)^b \psi ^c$ , $\Delta = (\frac{1728}{2} \psi ^6 +
\phi )^2 -1$ , where $ a=-\frac{1}{6},b=-\frac{2}{3}, c=1 $ and $\omega_0$ the
fundamental period (note that $\omega _0^2 | _{y=0}=E_4$).
 Now at weak coupling we have $\frac{\partial (x,y)}{\partial (q_1,q_2)}
\sim\frac{j_q g}{j^2} (q_1)$ , so that
%---------------------------------------------------------
\begin{eqnarray*}
\frac{\partial (\psi,\phi )}{\partial (t_1, t_2 )} \sim y^{-1} \psi^{-5}
j_{\tau} ,
\end{eqnarray*}
%---------------------------------------------------------

 which together with (remember $j=\frac{E_4^3}{\eta^{24}}$)

%---------------------------------------------------------
\begin{eqnarray*}
\psi &\sim& (\frac{1}{yx^2})^{\frac{1}{12}}=y^{-\frac{1}{12}}
\frac{E_4^{\frac{1}{2}}}{\eta^4}\\
\Delta|_{y\approx 0} &\sim& y^{-1} (j-j(i))^2
\end{eqnarray*}
%---------------------------------------------------------

 and with the relation $j_{\tau}^2 \sim j(j-j(i))E_4$ leads to the announced
expression of $F_1^{II}$ with

%---------------------------------------------------------
\begin{eqnarray*}
\alpha &=& 1+a+b+\frac{1}{12}(c-\frac{\chi}{12})\\
\beta  &=& 2a+\frac{1}{2}\\
\gamma &=& 2(c+3-\frac{\chi}{12})\;.
\end{eqnarray*}
%---------------------------------------------------------

 Now one actually has $c=1$ [\ref{CdlOFKM}]. Then , on the one hand, insertion
of the values $a=-\frac{1}{6}, b=-\frac{2}{3}$ obtained by comparing at the
large radius limit with the topological intersection numbers [\ref{CdlOFKM}]
leads  immediately to the correct {\em explicit} powers of $\alpha,
\beta,\gamma$. On the other hand one can also get the {\em abstract
cohomological meaning} of $\frac{\alpha}{\beta}$ and $\frac{\gamma}{\beta}$ on
the Calabi-Yau-side (in our two parameter model ) by noting the possibility of
direct comparison of the expression for $F_1^{II}$ with the large radius limit
(leaving aside the intermediate step of determination of a and b)  obtaining
the result

%---------------------------------------------------------
\begin{eqnarray*}
\alpha=\frac{1}{12} c_2\cdot L=\frac{1}{12} c_2(L)=\frac{1}{12} \chi _{K3}
\end{eqnarray*}
%---------------------------------------------------------

and

%---------------------------------------------------------
\begin{eqnarray*}
\beta + \frac{\gamma}{12} &=& \frac{1}{12} c_2 \cdot H\\
&=& \frac{1}{12}[2\chi _{K3}+c_2(E)-E^3]\\
&=& \frac{1}{12}[2\chi _{K3}+2\chi _C-4\chi _C]\\
&=& \frac{1}{12} 2(\chi _{K3}-\chi _C) ,
\end{eqnarray*}
%---------------------------------------------------------

(where we used the relation $E^3=-8=4\chi_C$ between the singular curve C of
the Calabi-Yau and the ruled surface E of its pointwise resolution
[\ref{CdlOFKM}]), so $12\beta
+\gamma=2(4-\frac{\chi}{12})(12\frac{\beta}{\gamma}+1)$ equals\footnote{Note
that because of the $K_3$-fibration $\chi =(\chi _{P1}-12)\chi _{K3}+12(\chi _C
+1)$ so $-\frac{\chi}{12}=-4+\chi _{K_3}-\chi _C -1$ ( the $+1$ comes from the
possibility of having $z_3=z_4=z_5=0$).} $2(\chi _{K_3}-\chi
_C)=2(4-\frac{\chi}{12}+1)$ from which it follows that
$2(4-\frac{\chi}{12})12\frac{\beta}{\gamma}=2\cdot1$, which shows that

%----------------------------------------------------------
\begin{eqnarray*}
\frac{\gamma}{\beta}&=& 12(4-\frac{\chi}{12})\\
\frac{\alpha}{\beta}&=&\frac{\chi
_{K3}}{12}\frac{12(4-\frac{\chi}{12})}{\gamma}=\frac{\chi _ {K3}}{2}.
\end{eqnarray*}
%----------------------------------------------------------

This agrees with the meaning of the numbers on the heterotic side.

%-----------------------------------------------------------

\hspace{0.5cm}

{\bf Acknowledgement:} I would like to thank  E. Derrick, G. Lopes-Cardoso , D.
L\"ust for  discussions.

\hspace{0.5cm}

%---------------------------------------------------------------------
%
%                        BIBLIOGRAPHY
%
%----------------------------------------------------------------------

\section*{References}
\begin{enumerate}
\item
\label{D}
M. Duff, Nucl. Phys. {\bf B442} (1995) 47, hep-th/9501030.

\item
\label{W}
E. Witten, Nucl. Phys. {\bf B443} (1995) 85, hep-th/9503124.

\item
\label{FHSV}
S. Ferrara, J. Harvey, A. Strominger, C. Vafa, hep-th/9505162.

\item
\label{KV}
S. Kachru, C. Vafa, hep-th/9505105.

\item
\label{KLM}
A. Klemm, W. Lerche, P. Mayr, hep-th/9506112.

\item
\label{KKLMV}
S. Kachru, A. KLemm, W. Lerche, P. Mayr, C. Vafa, hep-th/9508155

\item
\label{AGNT}
I. Antoniadis, E. Gava, K.S. Narain, T.R. Taylor, hep-th/9507115.

\item
\label{AP}
I. Antoniadis, H. Partouche, hep-th/9509009.

\item
\label{KLTh}
V. Kaplunovsky, J. Louis, S. Theisen, hep-th/9506110.

\item
\label{dWKLL}
B. de Wit, V. Kaplunovsky, J. Louis, D. L\"ust, hep-th/9504006.

\item
\label{BCOV}
M. Bershadsky, S. Cecotti, H. Ooguri, C. Vafa, Nucl. Phys. {\bf B405} (1993)
279; Comm. Math. Phys. {\bf 165} (1994) 311.

\item
\label{CdlOFKM}
P. Candelas, X.C. de la Ossa, A. Font, S. Katz, D. Morrison, Nucl. Phys. {\bf
B416} (1994) 481.

\end{enumerate}

%-----------------------------------------------------------------------------
\end{document}